\documentstyle[epsf,twocolumn,prl,aps]{revtex}

\begin{document}

\title{Finding the Pion in the Chiral Random Matrix Vacuum}
\author{G. W. Carter and A. D. Jackson \\
\small The Niels Bohr Institute, Blegdamsvej 17\\
\small DK--2100 Copenhagen \O, Denmark\\ }
\date{\today}
\maketitle

\begin{abstract}
The existence of a Goldstone boson is demonstrated in chiral random matrix 
theory.  After determining the effective coupling and calculating the scalar 
and pseudoscalar propagators, a random phase approximation summation reveals 
the massless pion and massive sigma modes expected whenever chiral symmetry 
is spontaneously broken.
\\[1.5mm]
\end{abstract}
\vspace{3mm}

Increasing evidence supports the applicability of chiral random matrix theory 
($\chi$RMT) \cite{SV} for describing the spontaneous breaking of chiral 
symmetry in quantum chromodynamics (QCD).  (For a review, see 
Ref.\,\cite{review}.)   Using the Banks-Casher relation \cite{BC}, the 
vacuum chiral condensate, $\Sigma$, can be related to the spectral density 
of the Dirac operator in the vicinity of zero eigenvalue.  This model has 
been extended to describe chiral symmetry restoration, for both finite 
temperature and baryon chemical potential \cite{HJSSV}, in what amounts 
to a mean field treatment.  Although these efforts have led to a schematic 
phase diagram for QCD as a function of $T$, $\mu$, and the current quark mass, 
$m$, other familiar vacuum phenomena associated with chiral 
symmetry breaking remain to be obtained.  In this letter, we consider 
quark-antiquark correlation functions in order to recover the Goldstone mode.  
This is a ubiquitous feature associated with the spontaneous breaking of a 
continuous global symmetry which is expected whenever the symmetry group is 
of order greater than one.  In $\chi$RMT, SU$_L(N_f) \times $SU$_R(N_f)$ chiral 
symmetry is broken into SU$_V(N_f)$ as is believed to occur in simplified QCD 
with two massless flavors.  Along with the chiral condensate, 
$\langle\bar\psi\psi\rangle_0$, we should expect $\chi$RMT to provide us 
with massless pseudoscalar (i.e., the pion) and massive scalar (i.e., the 
sigma) excitations.   

In order to extract this feature from a model of chiral symmetry breaking
at the quark level, a resummation in the spirit of the random phase 
approximation (RPA) can be performed to construct the current-current 
correlation function in a given channel.  This is shown schematically in 
Fig.\,1.  Evidently, this procedure requires two model-dependent quantities: 
an effective interaction between quarks and a correlator describing the 
propagation of a quark-antiquark pair in the gluonic background.  If we 
denote the coupling as $g$ and the ``bare'' $q{\bar q}$ propagator as $\Pi_0$, 
the full or ``dressed'' correlation function, $\Pi$, can be written as an 
infinite sum.  Each term in this sum is an alternating string of coupling 
constants and bare propagators as indicated in Fig.\,1.  This geometric
series is readily written as:
\begin{equation}
\Pi = \frac{g}{1 - \Pi_0 g}\, .
\label{RPA}
\end{equation}
In principle, both $g$ and $\Pi_0$ can depend on the momentum carried by 
the quark-antiquark pair in a model-dependent manner.  We now turn to the 
particular case of $\chi$RMT.
\begin{figure}[b]
\setlength\epsfxsize{8.5cm} \centerline{\epsfbox{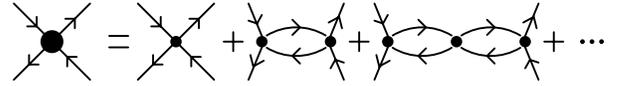}}
\caption{The $q{\bar q}$ propagator in the random phase approximation.}
\end{figure}

To determine $\Pi$, we assume a factorization of the elements $g$ and 
$\Pi_0$, in which case its solution takes the RPA-like form of
Eq.\,(\ref{RPA}) and Fig. 1.
The coupling constant is computed as the average strength of the four-quark 
interaction induced by the random matrix background.  Next, 
$\Pi_0$ is schematically associated with the two-eigenvalue correlation
function, calculable in $\chi$RMT, and evaluated in the appropriate limits.  
This is the zero-dimensional analogue of a loop integral, in that it 
encodes the effects of the chiral condensate on a quark-antiquark pair.

The effective coupling, $g$, is easily determined by averaging over the 
random matrix elements in the partition function of $\chi$RMT in order
to generate an effective quark interaction.  This partition function can 
be written as
\begin{eqnarray}
{\cal Z} &=& \int {\cal D}\psi^\dagger {\cal D}\psi {\cal D}W \;
{\rm exp}\Bigg\{ \sum_{j=1}^{N_f} \psi_j^\dagger \left(\begin{array}{cc} 
0 & iW \\ iW^\dagger & 0 \end{array} \right) \psi^j \nonumber\\
&&\qquad - n\Sigma^2 {\rm Tr}\left[ W^\dagger W\right] \Bigg\},
\label{PF}
\end{eqnarray}
where the $W$ are $n\times n$ complex matrices and where $\psi$ and 
$\psi^\dagger$ are row and column vectors with $2n$ components representing 
two-component chiral spinors.  Implicit in this form is the choice of a 
Gaussian unitary ensemble (GUE), appropriate for the SU(3) gauge group, with 
variance $\Sigma$.  The choice of a Gaussian potential is not mandatory 
in chiral random matrix theory since the microscopic spectral correlators have 
been shown to be universal for a wide class of potentials \cite{JSV,PHD}.  

The integral over $W$ in Eq.\,(\ref{PF}) can be performed to obtain
\begin{eqnarray}
{\cal Z} &=& \int {\cal D}\psi^\dagger {\cal D}\psi \; \nonumber\\
&&{\rm exp}\left(-\sum_{i,j=1}^{N_f}\;\sum_{\alpha,\beta=1}^n 
\frac{1}{n\Sigma^2} \psi_{Li\alpha}^\dagger\psi_R^{i\beta} 
\psi_{Rj\beta}^\dagger\psi_L^{j\alpha} \right),
\label{vertex}
\end{eqnarray}
from which one immediately finds an induced four-quark interaction.  After 
normalizing for $N$, the effective coupling strength is 
$g = (2 N_f \Sigma^2)^{-1}$.  
This coupling is constant, as is to be expected from the mean-field nature
of random matrix theory.

The second ingredient, the bare current-current correlation function, has 
the form 
\begin{eqnarray}
{\Pi_0}^\Gamma(q^2) &=& \int d^4x \; {\rm e}^{-iq \cdot x}
\left\langle \bar\psi(x) \Gamma \psi(x)
\; \bar\psi(0)\Gamma\psi(0) \right\rangle \nonumber\\
&=& \int\frac{d^4p}{(2\pi)^4}{\rm Tr}\left[iS(p)\Gamma iS(p-q)\Gamma\right]\,.
\label{cccf}
\end{eqnarray}
Here, $S(q)$ is the fermion propagator, $\Gamma$ is one of the 16
Dirac basis matrices which generate the Clifford algebra, and $q$ is 
the momentum carried by the quark-antiquark pair.  The two relevant 
correlators here are the identity matrix $\Gamma = {\bf 1}$ and $\Gamma = 
\gamma_5$ appropriate for the scalar and pseudoscalar channels, 
respectively.  We will be concerned only with results valid in
the $q\rightarrow 0$ limit of zero momentum transfer. 

Fortunately, the bare propagators of Eq.\,(\ref{cccf}) are directly related 
to the function, $\rho( \lambda_1 , \lambda_2 )$, describing the correlation 
between two eigenvalues of the Dirac operator.  This spectral correlator 
is accessible in $\chi$RMT and, moreover, known to be universal in the
microscopic limit.  The 
calculation of spectral correlators is usually technically involved, but 
it simplifies significantly in random matrix theory through the use of 
orthogonal polynomial techniques \cite{Metha}.  The desired spectral 
correlator is 
\begin{equation}
\rho(\lambda_1,\lambda_2) = \left\langle 
\frac{1}{N}\, {\rm Tr} \frac{1} {\lambda_1 - H}
\;\frac{1}{N}\, {\rm Tr} \frac{1} {\lambda_2 - H} \right\rangle
\label{rmcf}
\end{equation}
for eigenvalues $\lambda_1$ and $\lambda_2$ generated by averaging over
$N\times N$ matrices $H$ with elements drawn from a random Gaussian
distribution.  For problems with chiral symmetry, $H$ has the form
\begin{displaymath}
H = \left( \begin{array}{cc} 0 & iW \\ iW^\dagger & 0 \end{array}\right) \, .
\end{displaymath}

The scalar and pseudoscalar correlation functions of Eq.\,(\ref{cccf}) can 
be constructed from this spectral correlator.  We will be interested in 
taking limits $N\rightarrow \infty$ and $\Delta \equiv (\lambda_1-\lambda_2) 
\rightarrow 0$ simultaneously.  This requires first fixing $\Delta \sim 1/N$ 
and then taking $N \rightarrow \infty$.  
Since the effective coupling of Eq.\,(\ref{vertex}) does not depend on the 
momentum transferred, we are interested ultimately in the limit in which 
$\lambda\rightarrow 0$.  The limits cannot
be interchanged, for their ordering is crucial in obtaining the physically 
relevant quantities from $\chi$RMT \cite{JV}.

To extract the $\Delta$ dependence in the described limit we use the
microscopic spectral correlation function,
\begin{equation}
\rho_M(x,y) = \lim_{N\rightarrow\infty} \rho(\;x/N\Sigma\;,\;y/N\Sigma\;) \, .
\end{equation}
The microscopic limit is taken in order to retain information on the scale 
of the spacing between neighboring eigenvalues, the scale of $\delta \equiv
\Delta N \Sigma$ which corresponds to the momentum $q$.
At the same time the eigenvalues themselves are rescaled into 
microscopic variables $x = \lambda N \Sigma$ and the limit
$\lambda\rightarrow 0$ is replaced by $x \gg 1$.  
Thus, we are considering two eigenvalues which are close on the macroscopic
scale, but have a difference much greater than the average spacing between
microscopic, unfolded eigenvalues.

In these limits the real parts of the traces of Eq.\,(\ref{rmcf}) vanish,
being odd in $\lambda$, and we need only consider the contributions from
the imaginary parts.  These may be found by taking the discontinuities
across the real axes in the limit of zero current masses $m$, $m'$.
For a single fermion species,
\begin{eqnarray}
\lim_{q^2\rightarrow 0} \Pi(q^2) &=& \lim_{q^2\rightarrow 0}
\lim_{m,m'\rightarrow 0} {\rm Disc}\vert_{m=i\lambda,m'=i\lambda'}
\Pi(q^2;m,m') \nonumber\\
&=& \pi^2\lim_{\delta\rightarrow 0}\lim_{x\gg 1} 
\rho_M(x,x-\delta) \, .
\end{eqnarray}
Knowing that $\gamma_5$ reflects
left-handed quark eigenfunctions and leaves right-handed eigenfunctions 
unchanged, we find the bare scalar and pseudoscalar correlation functions 
in the appropriate limits:
\begin{eqnarray}
\lim_{q^2\rightarrow 0}\,{\Pi_0}^{{\rm S}}(q^2) &=&
2 N_f \pi^2 \lim_{\delta\rightarrow 0} \lim_{x\gg 1} 
\rho_M(x,x-\delta)\,, \nonumber\\
\lim_{q^2\rightarrow 0}\,{\Pi_0}^{{\rm PS}}(q^2) &=&
-2 N_f \pi^2 \lim_{\delta\rightarrow 0} \lim_{x\gg 1} 
\rho_M(x,x-\delta)\,.
\label{corrdefs}
\end{eqnarray}

The exact expression for the microscopic two point
correlation function, $\rho_M(x,y)$, was found 
in Ref.\,\cite{VZ} for chiral ensembles.  The connected part is
\begin{eqnarray}
\rho_M(x,y) &=& \frac{ \bar{x}\bar{y} }{ (x^2-y^2)^2 } \big[
x J_{N_f}(\bar{x}) J_{N_f-1}(\bar{y}) \nonumber\\
&&\qquad - y J_{N_f-1}(\bar{x}) J_{N_f}(\bar{y})\big]^2  \, ,
\label{twopoint}
\end{eqnarray}
where $\bar{x} = x \pi \rho(0) /\Sigma$, with $\rho(\lambda)$ being the
continuum single eigenvalue spectrum.

Fixing $\delta$ as previously explained, 
the above equation\,(\ref{twopoint}) has the limit
\begin{equation}
\lim_{x\gg 1}\rho_M(x,x-\delta) = \left( 
\frac{\sin[\delta\pi\rho(0)/\Sigma^2]}{\delta \pi/\Sigma^2} \right)^2\,.
\label{adj2}
\end{equation}

The distinction between the scalar and pseudoscalar correlators naturally 
lies in the sign difference between the two.
The argument $\delta$ is a small but finite number which can be 
associated with the Minkowski momentum, $q$, carried by the quark-antiquark 
pair.  Expanding Eq.\,(\ref{adj2}) in small $\delta$ and inserting this into
Eq.\,(\ref{corrdefs}), the quark correlation functions are found to be
\begin{equation}
{\Pi_0}^{\rm S,PS}(q^2) = \pm
2N_f \left( \frac{\pi^4\rho(0)^4}{3\Sigma^4} q^2 - \pi^2\rho(0)^2\right)
+ {\cal O}(q^4) 
\label{corrsolns}
\end{equation}
valid in the limit of small $q$.

Now, we need only perform the sum indicated in Fig.\,1.  The full correlation 
function, ${\Pi}^{\Gamma}(q^2)$, is summed as Eq.\,(\ref{RPA}):
\begin{equation}
{\Pi}^{\Gamma}(q^2) = \frac{1}{2N_f\Sigma^2 - {\Pi_0}^{\Gamma}(q^2)} \, .
\end{equation}
Inserting the solutions for each channel, Eqs.\,(\ref{corrsolns}), we find 
\begin{equation}
\Pi^S(q^2) = -\frac{C}{q^2 - 2 N_f C \left[\pi^2\rho(0)^2+\Sigma^2\right]},
\label{scprop}
\end{equation}
for the scalar channel, where $C = 3\Sigma^4/[2 N_f\pi^4\rho(0)^{4}]$.  For the 
pseudoscalar channel we obtain
\begin{equation}
\Pi^{PS}(q^2) = \frac{C}{q^2 - 2 N_f C \left[\pi^2\rho(0)^2 - \Sigma^2
\right]} \, .
\end{equation}
The scalar propagator has a pole at finite $q^2$.  This result is consistent 
with our expectation of a massive excitation.  (This is simply the scalar 
mesonic mode, usually denoted $\sigma$.)  On the other hand, we expect that 
pseudoscalar excitations should be identified with the pion and, hence, 
should be massless.  Indeed, by virtue of the Banks-Casher relation,
\begin{equation}
\Sigma = \pi\rho(0)\,,
\end{equation}
we see that the pseudoscalar propagator assumes the form 
\begin{equation}
\Pi^{PS}(q^2) = \frac{C}{q^2}\,,
\end{equation}
exhibiting the pole at $q^2=0$ indicative of a Goldstone boson.  This 
cancellation is precisely analogous to that which occurs when a gap equation,
arising in an effective field approach (e.g., the Nambu--Jona-Lasinio or 
instanton vacuum models), conspires with the $q$-independent part of the 
pseudoscalar correlator to leave a massless mode.  Here, $\chi$RMT exhibits 
a corresponding content in that a relation between the chiral condensate
and the single-quark energy spectrum is vital for recovering the 
pion as a massless excitation.  Furthermore, the collapse of support at
zero eigenvalue with the restoration of chiral symmetry \cite{JV} 
mimics the gap equation's loss of validity when the condensate vanishes.

Finally, we note that the scalar propagator of Eq.\,(\ref{scprop}) implies 
an effective mass of $M^2 = 6\Sigma^2$.  Because of the ambiguity of
scale in our association between eigenvalues and momenta this cannot be
taken as a precise numerical prediction.  However, it is clear that the
mass scales with the strength of the condensate, $\Sigma$, as is common
in effective field models.

The present results were obtained with Gaussian distributed 
random matrix elements.
However, the demonstration of massless pseudoscalar and massive scalar 
excitations relied only on properties of the microscopic spectral 
correlators which are known to be universal.  Thus, we expect that the 
present qualitative results will apply to a greatly extended class of 
potentials.  

By following the path of an RPA resummation and utilizing exact results from
random matrix theory in the thermodynamic limit, we have demonstrated that
evidence of spontaneously broken chiral symmetry based on the Banks-Casher 
relation is confirmed by the existence of Goldstone modes with mesonic 
quantum numbers.  This result indicates that $\chi$RMT inherently contains 
phenomena beyond the mere appearance of a chiral condensate.  It also suggests 
that the pattern of spontaneous symmetry breaking found in $\chi$RMT is the 
same as that occurring in dynamical field theoretic approaches.  Finally, the 
underlying connection between the vacuum condensate and the single-quark 
spectrum is further seen to be a feature which chiral random matrix theory 
shares with other descriptions of chiral symmetry breaking.  

We thank W. Weise for a particularly useful discussion, B. Vanderheyden
for numerous conversations, A. Polleri for a discussion of the NJL model, 
and K. Splittorff for a critical reading of this manuscript.

\end{document}